\documentclass[traditabstract]{aa} 

\usepackage{graphicx, times, txfonts}

\newcommand{\mm}{$\mu$m}
\newcommand{\lsol}{\mbox{L$_{\odot}$}}

\newcommand{\ks}{km s$^{-1}$}

\begin{document}

\title{An independent distance estimate to CW Leo\thanks{Herschel is
    an ESA space observatory with science instruments provided by
    European-led Principal Investigator consortia and with important
    participation from NASA.}  
}
 
\author{ 
M.A.T.~Groenewegen\inst{1}\and 
M.J.~Barlow\inst{2}\and
J.A.D.L.~Blommaert\inst{3}\and
J.~Cernicharo\inst{4}\and
L.~Decin\inst{3,5}\and
H.L.~Gomez\inst{6}\and
P.C.~Hargrave\inst{6}\and
F.~Kerschbaum\inst{7}\and
D.~Ladjal\inst{8}\and
T.L.~Lim\inst{9}\and
M.~Matsuura\inst{2}\and
G.~Olofsson\inst{10}\and
B.~Sibthorpe\inst{11}\and
B.M.~Swinyard\inst{11}\and
T.~Ueta\inst{8} \and
J.~Yates\inst{2}
} 
 

\institute{ 
Koninklijke Sterrenwacht van Belgi\"e, Ringlaan 3, B--1180 Brussel, Belgium 
\and
Department of Physics and Astronomy, University College London, Gower Street, London WC1E 6BT, UK
\and
Instituut voor Sterrenkunde, Universiteit van Leuven, Celestijnenlaan 200D, B--3001 Leuven, Belgium 
\and
Astrophysics Dept, CAB (INTA-CSIC), Crta Ajalvir km 4, 28805 Torrejon de Ardoz, Madrid, Spain 
\and
Sterrenkundig Instituut Anton Pannekoek, University of Amsterdam, 
Kruislaan 403, NL--1098 Amsterdam, The Netherlands
\and
School of Physics and Astronomy, Cardiff University, 5 The Parade, Cardiff, Wales CF24 3YB, UK
\and
University of Vienna, Department of Astrophysics, T\"urkenschanzstrasse 17, A--1180 Wien, Austria 
\and
Dept. of Physics and Astronomy, University of Denver, Mail Stop 6900, Denver, CO 80208, USA  
\and
Space Science and Technology Department, Rutherford Appleton Laboratory, Oxfordshire, OX11 0QX, UK  
\and
Dept of Astronomy, Stockholm University, AlbaNova University Center, 
Roslagstullsbacken 21, 10691 Stockholm, Sweden             
\and
UK Astronomy Technology Centre, Royal Observatory Edinburgh, Blackford Hill, Edinburgh EH9 3HJ, UK 
} 
 
\date{received: 2012,  accepted: 2012} 
 
 
 

\abstract{CW Leo has been observed six times between October 2009 and June 2012 with the SPIRE instrument on board the \it Herschel \rm satellite.
Variability has been detected in the flux emitted by the central star with a period of 639 $\pm$ 4  days, 
in good agreement with determinations in the literature.
Variability is also detected in the bow shock around CW Leo that had previously been detected in the ultraviolet and Herschel PACS/SPIRE data.
Although difficult to prove directly, our working hypothesis is that this variability is directly related to that of the central star.
In this case, fitting a sine curve with the period fixed to 639 days results in a time-lag in the variability 
between bow shock and the central star of 402 $\pm$ 37 days.
The orientation of the bow shock relative to the plane of the sky is unknown (but see below).
For an inclination angle of zero degrees, the observed time-lag translates into a distance to CW Leo of 130 $\pm$ 13 pc, and for non-zero inclination angles the distance is smaller.
Fitting the shape of the bow shock with an analytical model (Wilkin 1996), the effect of the inclination angle on the distance may be estimated.
Making the additional assumption that the relative peculiar velocity between the interstellar medium (ISM) and CW Leo is determined entirely by the star
space velocity with respect to the local standard of rest (i.e. a stationary ISM), the inclination angle is found to be 
 ($-$33.3 $\pm$ 0.8)\degr\ based on the observed proper motion and radial velocity.
Using the Wilkin model, our current best estimate of the distance to CW Leo is 123 $\pm$ 14 pc. 
For a distance of 123 pc, we derive a mean luminosity of 7790 $\pm$ 150 \lsol\  (internal error). 
}

\keywords{circumstellar matter -- infrared: stars -- stars: AGB and post-AGB -- stars: carbon  -- stars: individual: CW Leo}

\maketitle

\section{Introduction} 

CW Leo (= IRC~+10~216 = AFGL~1381) was discovered by Becklin et al. (1969) 
in the pioneering \it Two-micron Sky Survey \rm as an extremely red object. 
It soon turned out to be a carbon star (Miller 1970, Herbig \& Zappala 1970) 
in an advanced stage of stellar evolution called the asymptotic giant branch (AGB). 
It is pulsating and surrounded by an optically thick dust shell and 
a large molecular circumstellar envelope (CSE).

In the near- and mid-infrared (IR) it is one of the brightest objects in the sky,
thus a typical target for any new instrument or telescope operating
from the infrared to the millimetre.
With the {\it Herschel} satellite (Pilbratt et al. 2010) two important discoveries have already been
published on CW Leo: 
the discovery of many high-temperature water lines that have shed new
light on the origin of water around carbon stars (Decin et al. 2010),
and the confirmation of a bow shock, produced by the interaction of the stellar wind with the interstellar medium (ISM), 
originally discovered by Sahai \& Chronopoulos (2010) in the ultraviolet with {\it Galex}, by Ladjal et al. (2010, hereafter L10). 
%

Although an important object, its distance is uncertain, which is reflected in the 
uncertain estimates of basic quantities such as the luminosity and mass-loss rate.
One of the most in-depth studies was conducted by Groenewegen et al. (1998), where
dust and molecular radiative-transfer models were used to fit
simultaneously the available photometric data, 
the {\it Low Resolution Spectrometer} spectrum taken by the {\it Infrared Astronomical Satellite}, 
near- and mid-IR interferometric observations, and CO J= 1-0 up to 6-5 molecular line emission data, 
available at that time.
The conclusion was that the distance must be in the range 110-135 pc
(corresponding to a luminosity of 10~000 \lsol\ to 15~000 \lsol), which was consistent with
the luminosity of 7~700 \lsol\ to 12~500 \lsol\ based on the Mira
period-luminosity (PL-) relation (Groenewegen \& Whitelock 1996), taking
into account the scatter in that relation. 
Other distances quoted in the literature are based on slightly different versions of the  PL-relation, 
e.g. 120 pc (Scho\"ier et al. 2007) or 140 pc (Menzies et al. 2006).
%

In this work, an independent distance estimate to CW Leo is provided, based on the phase-lag between the flux variations of the central star and the bow shock.
In Section~2, the observations are presented, and the analysis is described in Section~3.
The model that was used to correct for the inclination angle of the bow shock is outlined in the Appendix.


\section{Observations} 

Imaging observations on board the \it Herschel \rm satellite with the SPIRE (Griffin et al. 2010) istrument 
have been taken on six separate occasions (see Table~\ref{TabObs}).
The observations were conducted in October 2009 and 2010 as part of the MESS guaranteed-time key program (Groenewegen et al. 2011), 
the observation in November 2009 was taken in SPIRE performance verification (PV) time and is publicly 
available through the Herschel science archive, and the last three observations were part of a 
DDT program (program DDT\_mgroen01\_6) with the specific aim of studying the variability of CW Leo. 
In all cases, the "Astronomical Observation Request" was identical, a SPIRE "Large Map" with a
repetition factor of 3. The map taken in PV has a size 4\arcmin\ x 4\arcmin\ and does not include the bow shock, 
while the other maps are 30\arcmin\ x 30\arcmin.

The image taken in October 2009, together with a complementary PACS (Poglitsch et al. 2010) image, 
was discussed in L10 and confirmed the presence of a bow shock around CW Leo 
that had been discovered by Sahai \& Chronopoulos (2010) in GALEX data.

\begin{figure} 

\includegraphics[angle=-0,width=80mm]{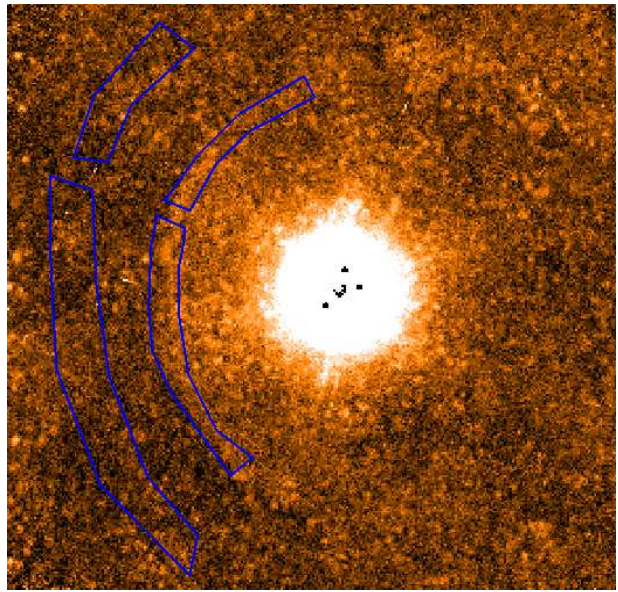}

\caption[]{ 
SPIRE PSW image with background sources removed, illustrating the location of the apertures used on the bow shock and the sky.
The size of the image is approximately 18 arcminutes on a side.
Some artifacts of the source removal may be seen close to the central object.
} 
\label{fig-contour} 
\end{figure}

The SPIRE data were reduced in a standard way (see Swinyard et al. 2010, Groenewegen et al. 2011) using the 
\it Herschel \rm Interactive Processing Environment (Ott et al. 2010, HIPE) version 8.2.0 in June 2012.
Aperture fluxes for the central object were determined using the {\it annularSkyAperturePhotometry} 
tool in HIPE (see Groenewegen et al. 2011), and are reported in Table~\ref{TabObs}.  
The apertures used were 237\arcsec, 196\arcsec, and 189\arcsec\ for the PSW (250 \mm), PMW (350 \mm), and  
PLW (500 \mm) filter, respectively.  
The beam areas to convert Jy/beam to Jy/pixel are 423, 751, and 1587 arcsec$^2$, respectively, in the three filters 
(SPIRE observers manual\footnote{http://herschel.esac.esa.int/Docs/SPIRE/html/spire\_om.html}) 
The fluxes for the first observation supersede 
those given in Groenewegen et al. (2011), which were calculated using the calibration files associated with 
HIPE 4.4.0\footnote{and were 138, 57.7, 23.2 Jy, in apertures of 211, 325, and 276\arcsec, respectively, and adopting beam areas of 501, 943, and 1923 arcsec$^2$, respectively.}.
To determine the flux in the bow shock, a slightly different approach was taken.
Background sources were removed from the image (they are irrelevant to the flux determination of the central star), using both the 
{\it sourceExtractorDaophot} and {\it sourceExtractorSussextractor} tools within HIPE.
The apertures for both the location of the bow shock and the sky were selected manually, 
and are shown in Fig.\ref{fig-contour}. As one can see, there is a small gap in-between the two apertures.
This is due to a diffraction spike at this location in the PACS 70 $\mu$m image, 
and this region should be excluded from the calculation of the flux from the bow shock at that wavelength (see L10). 
This problem is irrelevant to the present paper but does allow us to combine different on-source apertures with different 
sky apertures to better calculate the error in the flux determination.
Using a different sky aperture or doing one more pass of the background source removal task results in an estimated systematic error of about 0.12 Jy.
The random error in the actual flux measurement is estimated to be 0.03 Jy.
The absolute flux calibration error is estimated to be 7\% (SPIRE observers manual).
The last column of Table~\ref{TabObs} gives the flux of the bow shock in the PSW filter.
The fluxes in the other two SPIRE filters are much lower and are not reported here. 
Given the error in the fluxes, these two filters do not add to the variability information that is of concern in the present paper.
Although the full widthat half maximum of the PSW beam is only about 18\arcsec\ (SPIRE observers manual), the central star is very bright, 
and may still contribute to the flux at the location of the bow shock.  To estimate this, the flux was
determined in the exact same apertures as shown in Fig.~\ref{fig-contour} but mirrored along the y-axis. 
The resulting flux is low (typically $-0.1$ Jy in the 4 observations) and consistent with zero within the errors.
This means that the effect of the central star on the measured flux of the bow shock is negligible.


\begin{table}[!ht]

\setlength{\tabcolsep}{1.0mm}

\caption{Aperture fluxes in the SPIRE filters on the central star and bow shock (last column). }

\begin{tabular}{rrrrrrrrrrrr}
\hline
\hline
Date\tablefootmark{a} & Julian Date & ObsId\tablefootmark{b} & PSW  & PMW  & PLW    & PSW     \\
           &          &         &  250 $\mu$m  & 350 $\mu$m & 500 $\mu$m & 250 $\mu$m &  \\
           &          &         &  (Jy) & (Jy) & (Jy)     & (Jy) \\
\hline
2009-10-25 & 2455129.6 & 186293 & 164.5 & 66.1 & 26.9 & 1.417 \\ %
2009-11-11 & 2455147.2 & 186943 & 165.4 & 65.7 & 27.6 &       \tablefootmark{c} \\
2010-10-24 & 2455494.3 & 207040 & 144.6 & 58.7 & 23.7 & 1.280 \\ %
2011-05-30 & 2455711.6 & 221902 & 152.5 & 61.3 & 25.4 & 1.189 \\ %
2011-10-23 & 2455858.2 & 231352 & 177.7 & 70.2 & 28.4 & 1.176 \\ %
2012-06-03 & 2456082.2 & 246623 & 153.4 & 61.9 & 25.0 & 1.189 \\ %

\hline
\end{tabular}
\tablefoot{
\tablefoottext{a}{Format: yyyy-mm-dd.}
\tablefoottext{b}{Add 1342000000 to get the Observation ID.}
\tablefoottext{c}{This observation was taken in PV and the area covered did not include the bow shock}
}
\label{TabObs}
\vspace{-4mm}
\end{table}

\section{Results and discussion}

A sine curve of the form $F(t) = F_0 + A \cdot \sin \left( 2\, \pi\, (t-T_0)/P  \right)$ was fitted to the data, 
using the program Period04 (Lenz \& Breger 2005). The Monte Carlo option was used to estimate the error bars.
For the PSW, PMW, and PLW filters separately, periods of, respectively, $P$ = 635 $\pm$ 4, 638 $\pm$ 10, and 646 $\pm$ 4 days are found.
Independently, the PMW and PLW fluxes were scaled to the average level of the PSW flux, and the period was determined for the combined data
set of 18 points to give a period of 639 $\pm$ 4 days, which is consistent with the values above.
The derived period compares well to other determinations in the literature: Le Bertre (1992) presented lightcurves in many bands 
in the near- and mid-IR, and found an overall best-fit period of 649 days (no error bar given), and periods based on $K$-band lightcurves of 
644 $\pm$ 17 days (Witteborn et al. 1980),  636 $\pm$ 3 days (quoted in Ridgway \& Keady 1988, based on unpublished material), 
and 638 days (Dyck et al. 1991, no error bar given).

\begin{figure}
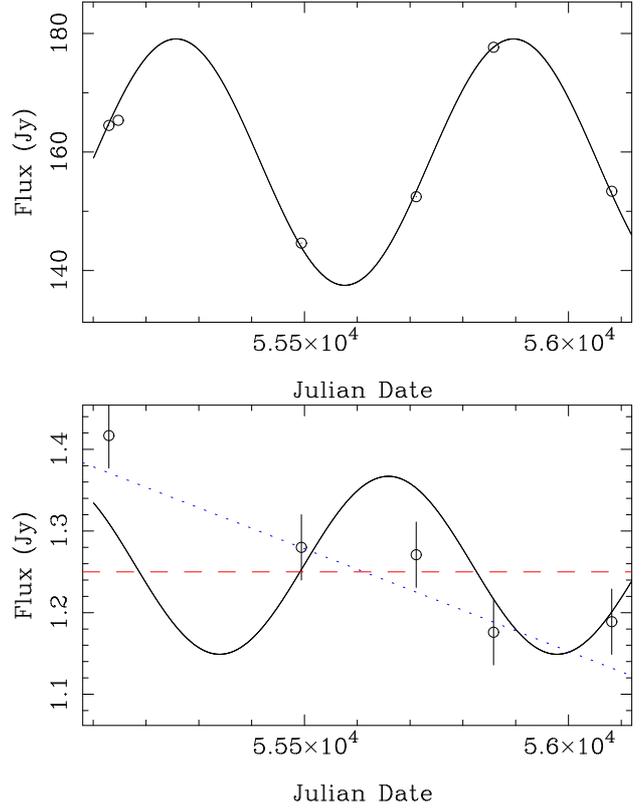
 


\includegraphics[angle=-0,width=82mm]{lcSTAR.ps}

\includegraphics[angle=-0,width=82mm]{lcBS.ps}

\caption[]{ 
Observations and fitted sinusoidal curve to the SPIRE PSW 250 \mm\ data on the central star (top panel), and the bow shock (lower panel).
The bottom panel also includes the best fit to the data using a constant (the red dashed line), and a line (the blue dotted line)
} 
\label{FigStar} 
\end{figure}

For the period fixed to 639 days, the amplitude, $A$, and zero level, $F_0$, of the sine curve were determined, and are listed in Table~\ref{TabRes}.
The time of maximum light is $T_o$ = 2455256.8 $\pm$ 2.2 for the lightcurve of the central star.
Taking the working hypothesis that the flux variation of the bow shock follows that of the central star, 
the time lag between the maximum light on the bow shock and the central star was determined to be 402 $\pm$ 37 days.
Figure~\ref{FigStar} shows the lightcurve of the star and the bowshock in the PSW filter, and the model fit to the observations.

The fit to the lightcurve of the bow shock flux is less secure,
and the five flux determinations may be equally well
fitted by either a constant or a line. 
A model with one parameter (a constant of 1.250, 
which is the average of the five determinations) results in $\chi^2$ of 24.0, 
and a value for the Bayesian information criterion\footnote{This
  is essentially a $\chi^2$ added with a term that penalizes models with
  more free parameters.} (Schwarz 1978) of BIC = 2.6. The sine model
(with three parameters, as the period is fixed) naturally results in
a lower $\chi^2$ of 13.0, but also in a lower BIC of -5.2.
The fit with a straight line (two parameters) formally fits the data best, with $\chi^2$ = 4.7, and BIC = -15.0.  
We do not have a plausible physical model that can explain why the flux on the bow shock would decrease exactly linearly with time.

\begin{figure} 

\includegraphics[angle=-0,width=80mm]{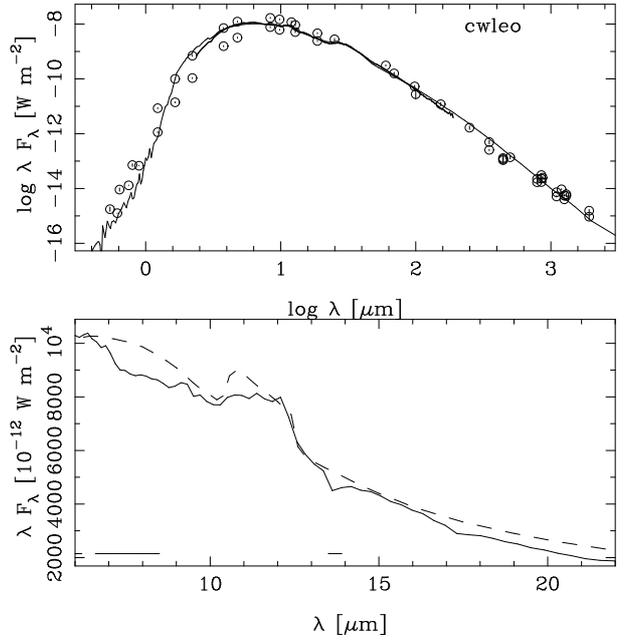}

\caption[]{ 
Fit to the SED (top panel), with a zoomed image of the 10 $\mu$m region in the lower panel.
 For a distance of 123 pc, the luminosity is 7790 \lsol.
The horizontal lines in the lower panel indicate the wavelength regions excluded from the fitting.
%
} 
\label{fig-sed} 
\vspace{-4mm}
\end{figure}

To illustrate our working hypothesis for the physical situation, the
dust radiative transfer model of Groenewegen (1997) for CW Leo was
updated with a newer version of the code (Groenewegen 2012), by
fitting the spectral energy distribution (SED), near- and mid-infrared
visibility curves, and PACS and SPIRE radial intensity profiles
(Groenewegen, in prep.).  The fit to the SED at mean light is
illustrated in Fig.~\ref{fig-sed}.  For the model shown in the figure,
the dust temperature at the location of the bow shock is 24.5~K.  This
is in excellent agreement with the fit to the PACS and SPIRE
photometry in L10, who derived 25 $\pm$ 3~K.  At that temperature, the
dust is primarily heated by photons emitted at about 110 $\mu$m.  The
optical depth at that wavelength is predicted to be 0.05, hence optically thin. 
The variation in flux of the central star and inner dust region would
therefore be felt directly at the location of the bow shock.

In a second model to represent the variation from minimum to maximum light, the effective temperature of the central star was increased by 300~K 
(see Men'shchikov et al. 2001) and the luminosity then increased as to reproduce the observed increase in SPIRE PSW flux of the central star.
In this model, the dust temperature at the location of the bow shock was increased from about 22.5~K to 27.5~K. 
This temperature variation alone would lead to a variation in flux of about 70\%, which is larger than is observed (20\%).
The exact change in the dust temperature depends however both on details of the model and the flux variation 
of the central star, which in turn depends on the details of the SPIRE calibration.

The model is also simplistic in the sense that the temperature calculated is that in the free expanding wind and 
not that in the bow shock, which is the result of the interaction of the wind with the interstellar medium. 
The calculation did show that the variation in theflux of the central star could lead to a variation in temperature 
and hence flux at the location of the bow shock, and that the flux variation is expected to follow the variation of the central star. 
The change in flux could also be due to a variation in dust density, 
but the timescale for the bow shock to adjust to changes in either the mass-loss rate of the central star 
(of order 500-1700 years, Decin et al. 2011) or the local density of the ISM is 
expected to be much longer than the pulsation period of the star (1.7 years).
Hydrodynamical models (van Marle et al. 2011, Cox et al. 2012) tuned to CW Leo may in the future lead 
to a better understanding of the nature of the dust emission.


\begin{table}[!ht]

\caption{Amplitude and zero level of the lightcurves in the SPIRE filters used to observe the central star and bow shock (Cols.5-7)}
\begin{tabular}{rrrrrrrrccc}
\hline
\hline
filter & $F_o$ &  $A$  & $A/F_o$ & $F_o$ &  $A$  & $A/F_o$ \\
       & (Jy)  & (Jy)  &         & (Jy)  & (Jy)  &          \\
\hline
PSW &  158.3 & 20.8  & 0.13 &  1.258 & 0.109 & 0.09 \\ 
PMW &   63.4 &  7.22 & 0.11 \\ 
PLW &   25.8 &  2.69 & 0.10 \\ 

\hline
\end{tabular}
\label{TabRes}
\vspace{-6mm}
\end{table}

The phase lag between the lightcurve measured on the central star and the bow shock allowed us to determine the distance to CW Leo.
The phase lag of (402 $\pm$ 37)  light days corresponds to (1.041 $\pm$ 0.096) 10$^{18}$ cm, and this translates to a relation 
between distance, $d$ (in pc), and angular separation ($\Delta \theta$) between the emission of the central star and the bow shock of 
$d \Delta \theta$ = (6.96 $\pm$ 0.64)  10$^{4}$ (\arcsec $\cdot$ pc). 

The distribution of the angular distance $\Delta \theta$ of all points inside the aperture shown in Fig.~\ref{fig-contour} 
and the central star was determined, and found to be (534 $\pm$ 16)\arcsec, based on the median value and 
the error estimated from half the difference between the 69\% and 31\% percentiles of the distribution.

If the bow shock were located in the plane-of-the-sky, the distance to CW Leo would follow immediately as $d$ = 130 $\pm$ 13 pc.
This is also an upper limit to the distance, as for bow shocks inclined with respect to the plane-of-the-sky the distance will be smaller.

To improve on this result, and obtain an estimate for the distance rather than just an upper limit, we employed a model 
that describes analytically the shape of a bow shock in the thin-shell limit (Wilkin 1996).
The model was used in L10 (also see Ueta et al. 2008, 2009).
In L10, we had assumed that the column density reaches its highest value where the bow shock cone intersects 
with the plane of the sky including the central star.
The Monte Carlo simulations of the three-dimensional (3-D) structure described in Appendix~A now show
that this is not the case, and that for non-zero inclinations of the
bow shock the surface brightness peaks at a location away from this plane (also see Cox et al. 2012).

Using these Monte Carlo simulations, it was possible to estimate for any
inclination the distribution of angular distances $\Delta \theta$ to the central star
of all points on the "Wilkinoid" that fall in the aperture when
projected on the sky. The results are listed in Table~\ref{TabIncl}, together with the distance that then follows.
We note that the fitting of the Wilkin model to the observed
trace of the bow shock in itself does not allow the inclination to be determined.
For zero inclination, the model gives a distance of (534.4 $\pm$ 18.3)\arcsec, in good agreement with the 
observed value of (534 $\pm$ 16)\arcsec.


\begin{table}[!ht]

\caption{Angular distance to the central star of all points on the "Wilkinoid" that fall in the aperture when
projected on the sky \rm for various inclination angles based on the Wilkin model, 
and the derived distance to CW Leo.}
\begin{tabular}{ccccclccc}
\hline
\hline
inclination &  $\Delta \theta$       & distance \\
 (\degr)    & (\arcsec)              & (pc)     \\
\hline

 0 & 534.4 $\pm$ 18.3 & 130.2 $\pm$ 12.7 \\ 
10 & 540.9 $\pm$ 19.8 & 128.6 $\pm$ 12.7 \\ 
20 & 550.5 $\pm$ 25.6 & 126.4 $\pm$ 12.9 \\ 
30 & 561.2 $\pm$ 30.4 & 124.0 $\pm$ 13.1 \\ 
33.3 & 564.7 $\pm$ 37.4 & 123.3 $\pm$ 13.7 \\ 
36 & 568.1 $\pm$ 39.0 & 122.5 $\pm$ 13.7 \\  
50 & 600.6 $\pm$ 56.0 & 115.9 $\pm$ 14.5 \\  
60 & 655.7 $\pm$ 77.0 & 106.1 $\pm$ 14.8 \\  

\hline
\end{tabular}
\label{TabIncl}
\vspace{-4mm}
\end{table}

Making one further assumption, we further refine our estimate for the distance to CW Leo.
The radial velocity of CW Leo is V$_{\rm LSR}= -25.5$ \ks\ (Groenewegen et al. 2002),
corresponding to V$_{\rm helio}= -18.6$ \ks, and its proper motion (PM) is
 $\mu_{\rm \alpha} \cos \delta = +35 \pm$1, $\mu_{\rm \delta} =+12 \pm$1 mas$\cdot$yr$^{-1}$ (Menten et al. 2012). 

At this point, we assume that the relative peculiar velocity between
the ISM and the star is determined entirely by the star
space velocity with respect to the local standard of rest (LSR)
(i.e. a stationary ISM). 
Then, following Cox et al., one can calculate the inclination angle.
Unfortunately, there is a typographical error in Table~1 of Cox et al. for CW Leo.

The correct values should read (for a distance of 123 pc, see below): 
a total PM of $\mu$ = 65 mas$\cdot$yr$^{-1}$, 
a peculiar space velocity of $v_{\star}$ = 45.7 \ks, 
position angle of PA = 66\degr, and, 
taking into account the errors in PM and RV (0.5 \ks\ adopted), an inclination angle of $i = -33.3$ $\pm$ 0.8\degr.

For this angle, one can take the true angular distance between the points located on the bow shock 
and the central star from Table~\ref{TabIncl}, and find our current best estimate of the distance to CW Leo of 123 $\pm$ 14 pc.

The model illustrated in  Fig.~\ref{fig-sed} leads to a luminosity of 7790 $\pm$ 150 \lsol\ for a distance of 123 pc.
Taking into account the error in the distance, we find $M_{\rm bol} = (-4.94 \pm 0.25$).
This is in agreement with the Mira PL-relation of Feast et al. (2006), 
$M_{\rm bol} = -2.54 \log P + 2.06 (\pm 0.24)$, which gives $(-5.07 \pm 0.24)$ for $P = 639$ days.
Menten et al. (2012) recently estimated the luminosity at phase 0.75 of the lightcurve (i.e. approximately mean light) 
from VLA observations to be $(8640 \pm 430)$ \lsol\ for 130 pc, or $(7730 \pm 380)$ \lsol\ for 123 pc, in excellent agreement with us.





\acknowledgements{  
%
JB, LD, and MG acknowledge support from the 
Belgian Federal Science Policy Office via the PRODEX Programme of ESA.
FK acknowledges funding by the Austrian Science Fund FWF under project number I163-N16 and P23586-N16.
SPIRE has been developed by a consortium of institutes led
by Cardiff Univ. (UK) and including: Univ. Lethbridge (Canada);
NAOC (China); CEA, LAM (France); IFSI, Univ. Padua (Italy);
IAC (Spain); Stockholm Observatory (Sweden); Imperial College
London, RAL, UCL-MSSL, UKATC, Univ. Sussex (UK); and Caltech,
JPL, NHSC, Univ. Colorado (USA). This development has been
supported by national funding agencies: CSA (Canada); NAOC
(China); CEA, CNES, CNRS (France); ASI (Italy); MCINN (Spain);
SNSB (Sweden); STFC, UKSA (UK); and NASA (USA).
%
}

{}

\begin{appendix}

\section{The Wilkin model}

In L10, the apparent shape of the bow shock was modelled following the
exact analytical solutions of Wilkin (1996), under certain assumptions.
In particular, we had assumed that the column density tends to reach its highest value 
where the bow shock cone intersects with the plane of the sky including the central star.
The Monte Carlo simulations of the 3-D structure described below show
that this is not the case, and that for non-zero inclinations of the
bow shock the surface brightness peaks at a location away from this plane.

\begin{figure} 

\includegraphics[angle=-0,width=85mm]{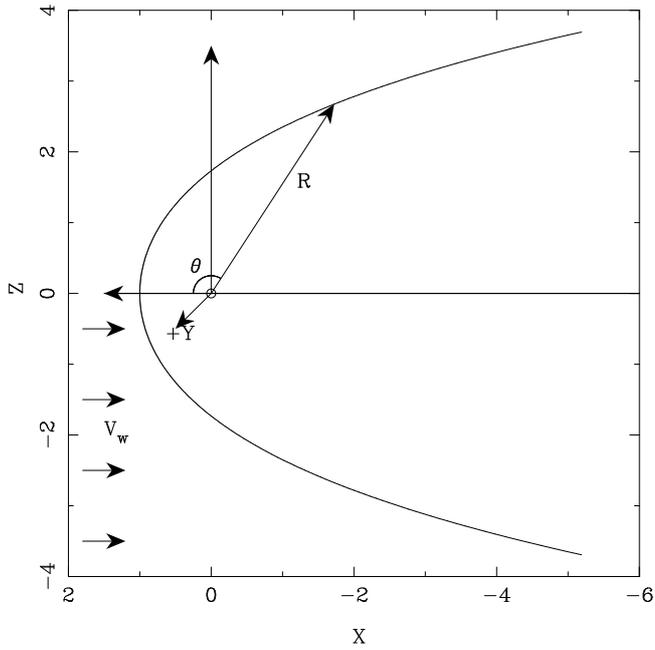}

\caption[]{ 
Definition of the right-handed coordinate system for the thin-shell bow shock model.
With reference to the plane of the sky, the positive x-axis points east, 
the positive y-axis points north, while the positive y-axis points towards the observer.
$\theta$ is the polar angle from the axis of symmetry, as seen from the star at the origin.
The azimuthal angle $\phi$ (not shown) is counted from the positive z-axis towards the positive y-axis.
The coordinate system may be rotated over the x-axis by an angle $\lambda$ counted in the same way as $\phi$,
over the y-axis by an angle PA (the position angle) counted from the positive x-axis towards the 
negative z-axis (i.e. south-of-east),
and over the z-axis by an angle $i$ (the inclination) counted positive from the positive x-axis 
towards the negative y-axis.
Shown is the Wilkin curve for a standoff distance of $R_0 = 1$.
The star is at rest and colliding head-on with a wind moving at a velocity $V_{\rm w}$.
} 
\label{fig-bsm} 
\end{figure}

Here we present a 3-D Monte Carlo simulation of the case where an
isotropic stellar wind interacts with the ISM of
homogeneous velocity $V_{\rm w}$ relative to the star and with a
stratified ISM density along the y-axis of the form $\rho = \rho_0 + a \, y$. 
This more complicated case than Wilkin (1996), where
$a = 0$, can also be described analytically (Wilkin 2000, and 
Canto et al. 2005; hereafter CRG). Here, we also assume thay $a = 0$.
The coordinate system is defined in Fig.~\ref{fig-bsm}.

The Monte Carlo simulation starts with drawing the angle $\theta$, $0 < \theta < \theta_{\rm max}$ 
($\theta_{\rm max}$ = 165\degr\ adopted) from a probability density function
\begin{equation} 
P(\theta) = \int_{0}^{\theta} \sigma(\theta^\prime) R(\theta^\prime) \sin (\theta^\prime) d \theta^\prime / P(\theta_{\rm max}), 
\label{Eq-P}
\end{equation} 
where  $\sigma$  is the mass surface density (Eq.~12 in Wilkin 1996).
The azimuthal angle  $\phi$ is a random value between 0 and $2\pi$.
$R(\theta, \phi$) can be solved from a third-order equation (Eq.~28 in CRG) for a given $\epsilon = a R_0 / \rho_0$, 
where $R_0$ is the so-called standoff distance.
For $a = 0$, $R$ is a function of $\theta$ only.
The velocities in the $z$ and $x$-direction are given by Eqs.~(17, 18, 33, 34, 35) in CRG.

The position and velocities in the cylindrical coordinate system are
then transformed to the $(x,y,z)$ system, which is then rotated over
specified angles $\lambda$, PA, and $i$ to the observers frame. The
outline of points can than be compared to the observed location of the bow shock, 
in order to infer the standoff distance, PA, and inclination (when $a = 0$ there is no dependence on the angle $\lambda$). 

\begin{figure} 

\includegraphics[angle=-0,width=50mm]{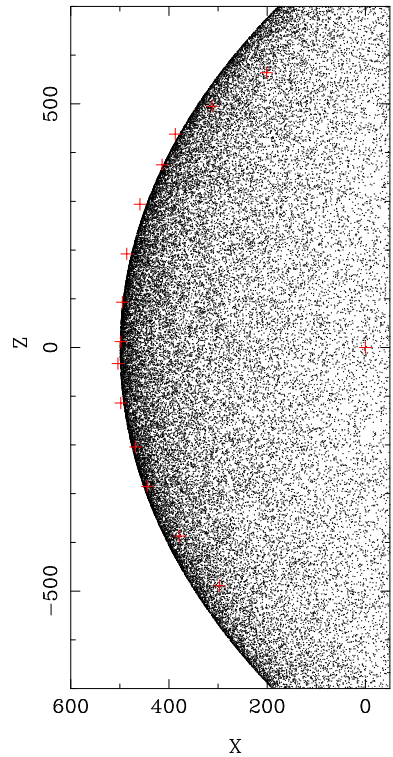}

\caption[]{ 
Monte Carlo simulation of a bow shock, for a standoff distance $R_0 = 499\arcsec$, 0\degr\ inclination, and position angle $-0.67\degr$.
CW Leo is at (0,0), and the units of the axis are in arcseconds.
The red crosses indicate the trace of the bow shock as seen with SPIRE at 250 $\mu$m (L10 and this paper).
} 
\label{fig-fit} 
\end{figure}

The results of the calculations are summarised in Table~\ref{TableW}, and an example of the fit to the observed trace is illustrated in Fig.~\ref{fig-fit}.
For a fixed inclination, the standoff distance $R_{\rm o}$ and position angle $PA$ were derived from a fit to the
trace of the bow shock in the SPIRE 250 $\mu$m filter (L10). The reduced $\chi^2$ ($\chi^2_{\rm r}$) is reported as a measure of the fit.
The reduced $\chi^2$ is quite large and  is related to the systematic deviation between observations and the Wilkin model for larger $Z$-values.
This probably indicates the limitations of the analytical model. We note that every simulated point is assumed to be equally 'observable'.
What is observed in reality is dust emission in the PSW filter, and the effect of changing the dust density and dust temperature along the bow shock is not considered here.
However, such effects are likely the reason why the bow shock can not be traced beyond $\sim \pm 500$\arcsec.
Since the procedure fits the trace of the bow shock, this should have little effect.

Although the smallest  $\chi^2$ are found for large inclination angles, the minimum is very shallow and the 
inclination angle cannot be derived from the Wilkin fitting (the same conclusion is reached by Cox et al. 2012).
The error quoted is the formal fit error. Monte Carlo simulations were performed allowing for a Gaussian error 
in the position of the trace of 3\arcsec\ (half a SPIRE PSW pixel) along the $z$-axis. The results show that the errors 
reported for $R_{\rm o}$ and PA are realistic, but also that the spread in the reduced $\chi^2$ is large, approximately 1 unit, 
indicating again that the inclinations angle cannot be derived from the Wilkin fitting alone.

For each combination of $i$, $R_{\rm o}$, and PA and 
every point inside the apertures shown in Fig.~\ref{fig-contour}, the true distance to the central star is recorded and are reported in Table~\ref{TabIncl}.

\begin{table}[!ht]
\setlength{\tabcolsep}{1.2mm}
\caption{Results of the Wilkin fitting.}
\begin{tabular}{cccccccc}
\hline
\hline
inclination & standoff distance & position angle  & reduced $\chi^2$ \\
   $i$      &   $R_{\rm o}$     &     PA (S-of-E)  & $\chi^2_{\rm r}$ \\
 (\degr  )  &   (arcsec)       &     (degrees)    & \\
\hline
 -0         & 499.1 $\pm$ 0.52 & $-0.67 \pm$ 0.49 & 20.9 \\ 
-10         & 493.2 $\pm$ 0.47 & $-0.46 \pm$ 0.43 & 20.9 \\
-20         & 478.7 $\pm$ 0.51 & $-0.35 \pm$ 0.48 & 20.5 \\ 
-30         & 453.9 $\pm$ 0.49 & $-0.50 \pm$ 0.43 & 19.8 \\ 

-33.3       & 443.4 $\pm$ 0.49 & $-0.12 \pm$ 0.43 & 19.6 \\

-36         & 434.3 $\pm$ 0.47 & $-0.25 \pm$ 0.43 & 19.4 \\

-45         & 398.5 $\pm$ 0.43 & $+0.08 \pm$ 0.37 & 18.4 \\ 


-50         & 375.6 $\pm$ 0.41 & $+0.03 \pm$ 0.33 & 17.8 \\ 
-60         & 322.3 $\pm$ 0.35 & $-0.03 \pm$ 0.25 & 16.4 \\
-70         & 258.8 $\pm$ 0.32 & $-0.11 \pm$ 0.20 & 14.7 \\ 

\hline
\end{tabular}
\label{TableW}
\end{table}

\end{appendix}

\end{document}